\input harvmac
\noblackbox
%%% Figures
\newcount\figno
\figno=0
\def\fig#1#2#3{
\par\begingroup\parindent=0pt\leftskip=1cm\rightskip=1cm\parindent=0pt
\baselineskip=11pt
\global\advance\figno by 1
\midinsert
\epsfxsize=#3
\centerline{\epsfbox{#2}}
\vskip 12pt
\centerline{{\bf Figure \the\figno:} #1}\par
\endinsert\endgroup\par}
\def\figlabel#1{\xdef#1{\the\figno}}

\def\np#1#2#3{Nucl. Phys. {\bf B#1} (#2) #3}
\def\pl#1#2#3{Phys. Lett. {\bf B#1} (#2) #3}
\def\prl#1#2#3{Phys. Rev. Lett. {\bf #1} (#2) #3}
\def\prd#1#2#3{Phys. Rev. {\bf D#1} (#2) #3}

%%% Paragraphs

%%% special math symbols
\font\cmss=cmss10
\font\cmsss=cmss10 at 7pt
\def\rlx{\relax\leavevmode}
\def\inbar{\vrule height1.5ex width.4pt depth0pt}
\def\IC{\relax\,\hbox{$\inbar\kern-.3em{\rm C}$}}
\def\IN{\relax{\rm I\kern-.18em N}}
\def\IP{\relax{\rm I\kern-.18em P}}
\def\ZZ{\rlx\leavevmode\ifmmode\mathchoice{\hbox{\cmss Z\kern-.4em Z}}
 {\hbox{\cmss Z\kern-.4em Z}}{\lower.9pt\hbox{\cmsss Z\kern-.36em Z}}
 {\lower1.2pt\hbox{\cmsss Z\kern-.36em Z}}\else{\cmss Z\kern-.4em
 Z}\fi}
%%% misc.
\def\IZ{\relax\ifmmode\mathchoice
{\hbox{\cmss Z\kern-.4em Z}}{\hbox{\cmss Z\kern-.4em Z}}
{\lower.9pt\hbox{\cmsss Z\kern-.4em Z}}
{\lower1.2pt\hbox{\cmsss Z\kern-.4em Z}}\else{\cmss Z\kern-.4em
Z}\fi}

\def\narrowplus{\kern -.04truein + \kern -.03truein}
\def\narrowminus{- \kern -.04truein}
\def\narrowminussub{\kern -.02truein - \kern -.01truein}

\def\g{{\gamma}}
\def\e{{\epsilon}}

\def\r{{\rightarrow}}

\def\frac#1#2{{#1\over #2}}

\def\com#1#2{{ \left[ #1, #2 \right] }}
\def\h#1#2{{ h_{#1}^{ \{ #2 \} }  }}
\def\acom#1#2{{ \left\{ #1, #2 \right\} }}
\def\xp{{ x\cdot p}}

\def\IZ{\relax\ifmmode\mathchoice
{\hbox{\cmss Z\kern-.4em Z}}{\hbox{\cmss Z\kern-.4em Z}}
{\lower.9pt\hbox{\cmsss Z\kern-.4em Z}}
{\lower1.2pt\hbox{\cmsss Z\kern-.4em Z}}\else{\cmss Z\kern-.4em
Z}\fi}
\def\IB{\relax{\rm I\kern-.18em B}}
\def\IC{{\relax\hbox{$\inbar\kern-.3em{\rm C}$}}}
\def\ID{\relax{\rm I\kern-.18em D}}
\def\IE{\relax{\rm I\kern-.18em E}}
\def\IF{\relax{\rm I\kern-.18em F}}
\def\IG{\relax\hbox{$\inbar\kern-.3em{\rm G}$}}
\def\IGa{\relax\hbox{${\rm I}\kern-.18em\Gamma$}}
\def\IH{\relax{\rm I\kern-.18em H}}
\def\II{\relax{\rm I\kern-.18em I}}
\def\IK{\relax{\rm I\kern-.18em K}}
\def\IP{\relax{\rm I\kern-.18em P}}
%\def\IX{\relax{\rm X\kern-.01em X}}
%this doesn't work

\def\p{\partial}

\font\cmss=cmss10 \font\cmsss=cmss10 at 7pt
\def\IR{\relax{\rm I\kern-.18em R}}

\def\f{\psi}

\def\dd{ {\delta}}

\def\df{\dot{\psi}}

\def\RV{{R_\Vert}}
%

%
%       \eqn\label{a+b=c}	gives displayed equation, numbered
%				consecutively within sections.
%     \eqnn and \eqna define labels in advance (of eqalign?)
%
\def\eqnn#1{\xdef #1{(\secsym\the\meqno)}\writedef{#1\leftbracket#1}%
\global\advance\meqno by1\wrlabeL#1}
\def\eqna#1{\xdef #1##1{\hbox{$(\secsym\the\meqno##1)$}}
\writedef{#1\numbersign1\leftbracket#1{\numbersign1}}%
\global\advance\meqno by1\wrlabeL{#1$\{\}$}}
\def\eqn#1#2{\xdef #1{(\secsym\the\meqno)}\writedef{#1\leftbracket#1}%
\global\advance\meqno by1$$#2\eqno#1\eqlabeL#1$$}

%%

% References for Non-Renorm

\lref\rK{N. Ishibashi, H. Kawai, Y. Kitazawa and A. Tsuchiya, hep-th/9612115.}
\lref\rCallias{C. Callias, Commun. Math. Phys. {\bf 62} (1978), 213.}
\lref\rPD{J. Polchinski, hep-th/9510017, \prl{\bf 75}{1995}{47}.}
\lref\rWDB{E. Witten,  hep-th/9510135, Nucl. Phys. {\bf B460} (1996) 335.}
\lref\rSSZ{S. Sethi, M. Stern, and E. Zaslow, Nucl. Phys. {\bf B457} (1995)
484.}
\lref\rGH{J. Gauntlett and J. Harvey, Nucl. Phys. {\bf B463} 287. }
\lref\rAS{A. Sen, Phys. Rev. {\bf D53} (1996) 2874; Phys. Rev. {\bf D54} (1996)
2964.}
\lref\rWI{E. Witten, Nucl. Phys. {\bf B202} (1982) 253.}
\lref\rPKT{P. K. Townsend, Phys. Lett. {\bf B350} (1995) 184.}
\lref\rWSD{E. Witten, Nucl. Phys. {\bf B443} (1995) 85.}
\lref\rASS{A. Strominger, Nucl. Phys. {\bf B451} (1995) 96.}
\lref\rBSV{M. Bershadsky, V. Sadov, and C. Vafa, Nucl. Phys. {\bf B463}
(1996) 420.}
\lref\rBSS{L. Brink, J. H. Schwarz and J. Scherk, Nucl. Phys. {\bf B121}
(1977) 77.}
\lref\rCH{M. Claudson and M. Halpern, Nucl. Phys. {\bf B250} (1985) 689.}
\lref\rSM{B. Simon, Ann. Phys. {\bf 146} (1983), 209.}
\lref\rGJ{J. Glimm and A. Jaffe, {\sl Quantum Physics, A Functional Integral
Point of View},
Springer-Verlag (New York), 1981.}
\lref\rADD{ U. H. Danielsson, G. Ferretti, B. Sundborg, Int. J. Mod. Phys. {\bf
A11} (1996) 5463\semi   D. Kabat and P. Pouliot, Phys. Rev. Lett. {\bf 77}
(1996), 1004.}
\lref\rDKPS{ M. R. Douglas, D. Kabat, P. Pouliot and S. Shenker,
hep-th/9608024,
Nucl. Phys. {\bf B485} (1997), 85.}
\lref\rhmon{S. Sethi and M. Stern, Phys. Lett. {\bf B398} (1997), 47.}
\lref\rBFSS{T. Banks, W. Fischler, S. H. Shenker, and L. Susskind,
Phys. Rev. {\bf D55} (1997) 5112.}
\lref\rBHN{ B. de Wit, J. Hoppe and H. Nicolai, Nucl. Phys. {\bf B305}
(1988), 545\semi
B. de Wit, M. M. Luscher, and H. Nicolai, Nucl. Phys. {\bf B320} (1989),
135\semi
B. de Wit, V. Marquard, and H. Nicolai, Comm. Math. Phys. {\bf 128} (1990),
39.}
\lref\rT{ P. Townsend, Phys. Lett. {\bf B373} (1996) 68.}
\lref\rLS{L. Susskind, hep-th/9704080.}
\lref\rFH{J. Frohlich and J. Hoppe, hep-th/9701119.}
\lref\rAg{S. Agmon, {\it Lectures on Exponential Decay of Solutions of
Second-Order Elliptic Equations}, Princeton University Press (Princeton) 1982.}
\lref\rY{P. Yi, hep-th/9704098.}
\lref\rDLhet{ D. Lowe, hep-th/9704041.}
\lref\rqm{M. Claudson and M. Halpern, \np{250}{1985}{689}\semi
R. Flume, Ann. Phys. {\bf 164} (1985) 189\semi
M. Baake, P. Reinecke and V. Rittenberg, J. Math. Phys. {\bf 26} (1985) 1070.}
\lref\rbb{K. Becker and M. Becker, hep-th/9705091, \np{506}{1997}{48}\semi
K. Becker, M. Becker, J. Polchinski and A. Tseytlin, hep-th/9706072,
\prd{56}{1997}{3174}.}
\lref\rss{S. Sethi and M. Stern, hep-th/9705046. }
\lref\rpw{J. Plefka and A. Waldron, hep-th/9710104, \np{512}{1998}{460}.}
\lref\rhs{M. Halpern and C. Schwartz, hep-th/9712133.}
\lref\rlimit{N. Seiberg hep-th/9710009, \prl{79}{1997}{3577}\semi
A. Sen, hep-th/9709220.}
\lref\rentin{D.-E. Diaconescu and R. Entin, hep-th/9706059,
\prd{56}{1997}{8045}.}
\lref\rgreen{M. B. Green and M. Gutperle, hep-th/9701093, \np{498}{1997}{195}.}
\lref\rpioline{B. Pioline, hep-th/9804023.}
\lref\rgl{O. Ganor and L. Motl, hep-th/9803108.}
\lref\rds{M. Dine and N. Seiberg, hep-th/9705057, \pl{409}{1997}{209}.}
\lref\rberg{E. Bergshoeff, M. Rakowski and E. Sezgin, \pl{185}{1987}{371}.}
\lref\rBHP{M. Barrio, R. Helling and G. Polhemus, hep-th/9801189.}
\lref\rper{P. Berglund and D. Minic, hep-th/9708063, \pl{415}{1997}{122}.}
\lref\rspin{P. Kraus, hep-th/9709199, \pl{419}{1998}{73}\semi
J. Harvey, hep-th/9706039\semi
J. Morales, C. Scrucca and M. Serone, hep-th/9709063, \pl{417}{1998}{233}.}
\lref\rdine{M. Dine, R. Echols and J. Gray, hep-th/9805007.}
\lref\rber{D. Berenstein and R. Corrado, hep-th/9702108, \pl{406}{1997}{37}.}

\Title{\vbox{\hbox{hep-th/9805018}
\hbox{DUK-CGTP-98-02, IASSNS--HEP--98/33}}}
{\vbox{\centerline{Constraints From Extended Supersymmetry}
\vskip8pt\centerline{in Quantum Mechanics}}}
%\smallskip
\centerline{Sonia Paban$^\ast$\footnote{$^1$} {paban@sns.ias.edu}, Savdeep
Sethi$^\ast$\footnote{$^2$} {sethi@sns.ias.edu} and Mark
Stern$^\dagger$\footnote{$^3$} {stern@math.duke.edu} }
\vskip 0.1in
\medskip\centerline{\it $^\ast$School of Natural Sciences}
\centerline{\it Institute for Advanced Study}\centerline{\it
Princeton, NJ
08540, USA}
%\vskip 0.05in
\centerline{and}
%\vskip 0.05in
\medskip\centerline{\it $^\dagger$Department of Mathematics}
\centerline{\it Duke University}\centerline{\it Durham, NC 27706, USA}

\vskip 0.25in

We consider quantum mechanical gauge theories with sixteen supersymmetries.
The Hamiltonians or Lagrangians characterizing these theories can contain
higher
derivative terms. In the operator approach, we show that the
free theory is
essentially the unique abelian theory with up to four derivatives in the
following sense: any small deformation of
the free theory, which preserves the
supersymmetries, can be gauged away by a unitary conjugation. We also present a
method for
deriving constraints on terms appearing in an effective Lagrangian.
We apply this method to the effective Lagrangian describing the
dynamics of two well-separated clusters of D0-branes. As a result, we prove a
non-renormalization theorem for the $v^4$ interaction.

%\draftmode
\vskip 0.1in
\Date{5/98}

\newsec{Introduction}

Quantum mechanical theories play an important role in recent
attempts to define M theory and field theories in various dimensions. One
of the more important systems is the  quantum mechanical gauge theory that
 describes the low-energy dynamics of zero-branes in type IIA string theory
\refs{\rPD, \rWDB}.
The system can be obtained by a
 dimensional reduction of supersymmetric Yang-Mills from ten dimensions \rqm.
The
theory has sixteen supersymmetries and a $U(N)$ gauge symmetry. BFSS have
conjectured that this matrix model describes M theory in eleven dimensions
in a limit where $N \r \infty$ \rBFSS. For finite $N$, the matrix model is
 believed to describe M theory quantized in the discrete light-cone formalism
(DLCQ) \refs{\rLS, \rlimit}.

The full gauge theories that appear in these matrix models are difficult to
study, particularly when $N$ becomes large. Fortunately, a key feature of these
gauge theories is the existence of flat directions on which scattering states
localize. For many of the questions that we might wish to answer, it is
sufficient
to control the physics of the light modes propagating along these flat
directions.
There are two distinct ways to go about analyzing the effective dynamics on the
flat directions: in an operator approach, we can use an integration procedure
of the
sort developed in \rss\ and further developed in \refs{\rpw, \rhs}. This
integration
procedure requires some knowledge of the bound state wavefunctions, which makes
it
difficult to extend to large $N$.

A second approach involves the perturbative construction of an effective
Lagrangian in
a velocity expansion. To date, there have been a number of computations of
loop corrections to the bosonic part of the effective
action \refs{\rDKPS, \rber, \rbb}, which takes the form:

\eqn\action{ \, S = \int{ dt \, \left( \, f_1(r) v^2
+ f_2(r) v^4 + \ldots \, \right).
}}
To order $v^2$, this theory can be canonically quantized so there is always an
effective Hamiltonian corresponding to this effective Lagrangian. This is
generally
not the case when higher velocity terms are included. To obtain a
supersymmetric
completion of \action, we typically need to add terms involving accelerations
and terms
with more than a single time derivative acting on a fermion. In this case, the
Lagrangian is not related to an unconstrained supersymmetric Hamiltonian
in any straightforward way.

The aim of this paper is to explore the extent to which supersymmetry
constrains
both the form of the Hamiltonian and the terms that can appear in the effective
Lagrangian \action. At order $v^2$, the two questions are closely related and
in the
Lagrangian approach, we need to
determine which metrics, specified by $f_1$, are compatible
with supersymmetry. We will show in section two that
supersymmetry actually constrains the metric to be flat.\foot{A fact often
stated
in the literature, but proven in no previous work of which we are aware.}

 Before going further, it is worth
clarifying some points. The low-energy description of a gauge theory differs
from a
conventional sigma model. In more conventional sigma models, the
fermions are sections of the tangent bundle, while in the case that we wish
to understand, the fermions transform as sections of the spin bundle. The
condition
that we have sixteen supersymmetries corresponds to the existence of sixteen
Dirac operators which act on the Hilbert space. We will take
our Dirac operators or supercharges $Q_a$ to be functions of bosons $x^i$ and
 fermions
$ \f_a$, where $i=1,\ldots, 9$ and $a=1,\ldots,16$. If we use $V$ to denote the
representation space of $Spin(9)$, which is a
sixteen-dimensional space, then our wavefunctions are spinors of the bundle
$V$.
Note that our conclusion about the metric does not
rule out the existence of a sigma model with a non-trivial metric,
sixteen supersymmetries, and a different fermion content.

For theories with
less supersymmetry, the metric is less constrained. With four supersymmetries,
the metric is determined by an arbitrary real function of $r$. With eight
supersymmetries, it was shown in an interesting paper \rentin\ using superspace
techniques that the only allowed metrics are harmonic functions in
five dimensions.

The next step is to consider terms to order $v^4$. In the operator approach
discussed in section three, we
will consider small deformations of the supercharges. We require that these
deformations
be compatible with a symmetry inherited from the full non-abelian gauge theory.
This symmetry is essentially CPT. To this order in a derivative expansion, we
will
show that any small
deformation of the supercharge can actually be gauged away by a unitary
transformation. Since we are interested physically
in interactions
that fall off sufficiently fast at infinity, it is enough to study small
deformations of the supercharge.  Our results then show that in this class of
theories, the free theory is
the unique gauge theory with sixteen supersymmetries.  It would be interesting
to
extend these results in two ways: first by
considering higher order terms in the derivative expansion and second by
considering
non-abelian gauge theories.

In the Lagrangian approach,
we know from loop computations that there are non-trivial higher order
interactions.
In section four, we examine constraints on the effective Lagrangian.
We wish to know what choices of $f_2(r)$ admit a supersymmetric completion. To
find a constraint, we study the eight fermion term in the supersymmetric
completion of $ f_2(r)\, v^4$. This leads to the constraint that $f_2$ must be
harmonic in nine dimensions which is the desired non-renormalization theorem.
This
ensures that the interaction between two gravitons in matrix theory agrees with
supergravity \rBFSS.

This method for finding constraints on the effective action can be extended to
both
higher velocity terms in the quantum mechanics and to field theories
in various dimensions. In four dimensions, a non-renormalization theorem for
the four derivative terms in the Yang-Mills effective action was proven in
\rds\
using arguments of a quite different flavor. Our results imply corresponding
non-renormalization theorems for all higher dimensional Yang-Mills theories,
except
three-dimensional Yang-Mills.\foot{Three-dimensional Yang-Mills is exceptional
because
of vector-scalar duality and will be studied elsewhere.} With sixteen
supersymmetries, it might well be the case that
there are some restrictions on which terms can appear at
every order in the velocity expansion. It would be very interesting to see
whether
this is the case. Finally, we should point out that our results have close
parallels with recent
restrictions on higher order corrections to effective actions in string theory
\refs{
\rgreen, \rpioline} and field theory \rgl. After we completed this project, an
interesting paper appeared with further evidence of a non-renormalization
theorem for the $v^4$ terms \rdine.

\newsec{Some Generalities}
\subsec{Constraining the metric}

The $Spin(9)$ Clifford algebra can be represented
by real symmetric matrices $\g^i_{ab}$, where $i=1,\ldots,9$ and
$a=1,\ldots,16$. These matrices obey:
\eqn\clifford{ \{ \g^i, \g^j \} = 2 \delta^{ij}. }
All the operators that we wish to study can be constructed in terms of the
basis $\left\{ I, \g^i, \g^{ij}, \g^{ijk}, \g^{ijkl} \right\}$, where we
define:
\eqn\defs{ \eqalign{ \g^{ij} &= {1\over 2!} ( \g^i \g^j - \g^j \g^i) \cr
\g^{ijk} &= {1\over 3!}( \g^i \g^j \g^k - \g^j \g^i \g^k + \ldots) \cr
\g^{ijkl} &= {1\over 4!}( \g^i \g^j \g^k \g^l - \g^j \g^i \g^k \g^l + \ldots).
 \cr}}
Note that $\left\{ I, \g^i, \g^{ijkl} \right\}$ are symmetric while
 $\left\{ \g^{ij}, \g^{ijk} \right\}$ are antisymmetric. The normalizations in
\defs\ are chosen so that the trace of the square
of a basis element is $ \pm 16$.

To fix the metric, it is easiest to work in the Lagrangian approach.
The most general $Spin(9)$ invariant metric takes the form:

\eqn\metric{ \eqalign{ ds^2 &= g_{\mu\nu} dx^{\mu} dx^{\nu}  = g_1(r) dx^i dx^i
+
g_2(r) x^i x^j dx^i dx^j, \cr & = f_1(r) dx^i dx^i} }
The last step in \metric\ is always possible by a coordinate choice. The
Lagrangian
is then,
\eqn\metriclag{L = \int{ dt \, f_1(r) v^2 + \ldots. }}
The supersymmetry transformations are highly constrained at this order. Since
this
system is a special case of a theory with $N=4$ supersymmetry which has a
superspace
formulation (at this order) \rentin, the supersymmetry transformations can only
be
a more restricted form of the $N=4$ transformations.  To be compatible with
$Spin(9)$ invariance, the supersymmetry transformations must take the form (at
this
order):
\eqn\transforms{ \eqalign{ \dd x^i & = -i \e \g^i \f \cr
\dd \f_a &= ( \g^i v^i \e )_a + ( M \e )_a.}}
The Grassmann parameter $\e$ is a $16$ component real spinor and $M$ is an
order $v$
expression containing two fermions.
To show the metric is flat, we need to show that $M$ must vanish. The algebra
must
close on time translations in the Lagrangian approach. Since,
\eqn\closure{\eqalign{[d_1, d_2] x^i &  = -i ( \e_2 \g^i (v^k \g^k + M) \e_1 -
\e_1 \g^i (v^k \g^k +M) \e_2 )\cr
& = -i \e_2 ( \{ \g^i , \g^k \} v^k + \{ \g^i M + M^T \g^i \} ) \e_1 \cr
& = -2i \e_2 \e_1 v^i -i \e_2 \{ \g^i M + M^T \g^i \} \e_1, }}
the second term containing $M$ must vanish. It
is not hard to check that
satisfying the condition $ ( \g^i M + M^T \g^i ) = 0$ requires $M$ to vanish.
Therefore the metric must be flat.

\subsec{The momentum expansion}

Let us now turn to the operator approach. As operators, the sixteen real
fermions
$ \f_a$ satisfy the anti-commutation relations,
\eqn\commutation{ \{ \f_a, \f_b \} = 2 \delta_{ab}. }
The supercharge $Q_a$
is hermitian with respect to the flat metric and must obey
the algebra,
\eqn\susyalgebra{ \left\{ Q_a, Q_b \right\} = 2 \delta_{ab} H,}
with no additional central terms on the right side. The requirement of a flat
metric
derived in the previous section can also be derived from the requirement that
the supercharge satisfy \susyalgebra, but the argument is more involved.

The algebra \susyalgebra\ is to be contrasted
with quantum mechanical gauge theories with charged fields which close on the
Hamiltonian
only up to gauge transformations. Obeying the algebra \susyalgebra\ provides a
strong
constraint on the allowed deformations of the charges.

The supercharge $Q_a$ can be expanded in the number of fermions,
\eqn\expansion{Q_a = f_{ac} \f_c + f_{acde} \f_c \f_d \f_e + \ldots,}
where $f_{acde}$ is antisymmetric in the last three indices etc. Let us
 introduce momenta conjugate to $x$ obeying the usual relation,
$$ \com{x^i}{p_j} = i\, \delta^{i}_j. $$
The basic $Spin(9)$ invariant combinations are $ \left( x^2, p^2, \xp \right)$.
Then the most general possible form for $f_{ac}$ is,
\eqn\lowestterm{ f_{ac} = \delta_{ac} D  + \g^i_{ac} D_i + \g^{ij}_{ac}
D_{ij},}
where we have specified nothing about the particular dependence of the
operators
$D, D_i, D_{ij}$ on the momenta. We can then express each of these operators
in terms of unknown functions of  $ \left( x^2, p^2, \xp \right)$:
\eqn\defh{ \eqalign{  D &= h_1 \cr  D_i &= h_2 p_i + h_3 x_i \cr
 D_{ij} &= h_4 (x_i p_j - x_j p_i). \cr}}
Now it should be clear why $\g^{ijk}$ did not appear in \lowestterm. That would
require the existence of an operator $D_{ijk}$ anti-symmetric in $i,j,k$ but
such an operator cannot be constructed.

It is actually easy to construct charges that satisfy the supersymmetry
algebra. For
example, we could simply set $D_i = D_{ij} = 0$ and take any hermitian $D$. The
resulting Hamiltonian $H=D^2$ contains no fermion terms. These are essentially
trivial solutions and we would like to rule out these cases. Actually, the
charges
that appear in physical contexts, such as the abelian theories describing the
low-energy
dynamics of quantum mechanical gauge theories, usually have $D=D_{ij}=0$. We
will not,
however, impose so strong a constraint quite yet. Rather, to rule out
uninteresting
solutions, we will require $h_2$ to be non-zero.

We will also eventually require that as $|x| \r
\infty$, the supercharge reduce to the charge $Q^0_a$ for a $U(1)$ gauge
theory:
\eqn\free{ Q^0_a = \g^i_{ac} \f_c p^i. }
This weak restriction is completely natural for most physical models, where
interactions
become weak as the distances become large.

The functions $h_i$ can be expanded in powers of $ p$. The usual counting
parameter is
the number of momenta plus half the number of fermions. This gives an expansion
in powers of
$ \hbar$ which we have generally set to one. For example to lowest order in
$\hbar$,
the supercharge is schematically,
$$ Q \sim \f p  + \f^3 + \f h, $$
where we have included the possibility of a static potential with the $h$ term.
Let us conveniently normalize our supercharges so that these lowest order terms
are
$O(\hbar)$. Then every additional momentum brings a power of $\hbar$ as do
every
two fermions. If there are non-trivial choices for some of the functions in
\defh\
then there will exist
 non-trivial solutions
for $Q$ which satisfy the supersymmetry algebra. What do we mean by a
non-trivial
solution?
At first sight, the answer seems self-evident, but that is actually not the
case.

\subsec{Gauge transformations}

We need to discuss a class of deformations of the supercharge that {\it do not
change} the physical system. The deformations correspond to unitary
transformations
of the supercharge,
\eqn\unitary{ Q_a \,\,\,\,\r \,\,\,\, e^{i C} Q_a e^{-i C},}
where $C$ is hermitian. Any conjugation of this kind automatically gives a new
set of
charges that satisfy the
supersymmetry algebra. To leading order in $C$,
 this transformation takes the form:
\eqn\leading{ Q_a \,\,\,\, \r \,\,\,\, Q_a + i  \left[ C, Q_a \right]. }
For example, the charge
\eqn\example{ Q_a = ( \g^i \f )_a  \left( p_i - p_j f_i p_j \right), }
is hermitian and obeys the algebra,
\eqn\closure{ \{ Q_a, Q_b \} = 2 \delta_{ab} (p^2 - 2 p_j f_i p_j p_i + i p_j
f_{ii} p_j)
+ \ldots,}
for any $f_i = \p_i f(r)$ and where the omitted terms are higher order in
${\hbar}$.
However, to leading order in $\hbar$, this
deformation can be written as a gauge
transformation where $C = p_j f p_j$ and so does
not result in a physically
distinct system. We should therefore only consider deformations modulo these
`gauge' transformations.

We can consider the possible $C$ operators both compatible with $Spin(9)$
invariance
and first order in  powers of momenta.
Up to first order, the only allowed operators are:
\eqn\first{ C = p_i k_i + k_i p_i + t(r).}
As usual, $k_i$ can written as $x_i k(r)$ for some radial function $k$. This
deformation seems to generate a `metric' at leading order \leading\ but the
rescaling of
the kinetic term is compensated by the induced scalar potential so the physical
metric
is unchanged. The key point is that we will meet
certain deformations like \example\ which appear to give new physical systems
where the
Hamiltonian has terms higher order in momenta, but
are actually gauge transformations. Our goal is to study deformations of the
supercharge which cannot be undone by some unitary transformation.

\newsec{Constraining the Supercharge}

\subsec{The terms leading in $\hbar$}

To study the leading order terms, we can truncate the expansion \expansion\ at
three fermions. Note that we will use the fact that the Hilbert space metric is
flat;
if this were not
the case then, for example, $p_i$ would not be hermitian. The three fermion
term cannot
appear with any momentum operators at
this order. We can then plug the charge into the supersymmetry algebra to
obtain,

\eqn\plugin{ \eqalign{ \left\{ Q_a, Q_b \right\} =&  \left\{ f_{ac},
f_{bc} \right \} -
 6 \acom{ f_{acde}}{ f_{bcde}}  \cr + &
 \left(  \com{f_{ak}}{ f_{bl}} + 3 \acom{f_{ac}}{ f_{bckl}}
+ 3 \acom{f_{bc}}{ f_{ackl}} - 18 \com{f_{amnk}}{f_{bmnl}} \right)
\f_{kl} \cr + & \left( \com{ f_{ak}}{f_{bmnl}} + \com{f_{bk}}{
f_{amnl}}
 + 9 \acom{f_{ackl}}{ f_{bcmn}} \right) \f_{klmn} \cr
+ & \com{f_{acde}}{f_{bmnr}} \f_{cdemnr}. \cr}}
 A summation over repeated indices is assumed and we have defined:

\eqn\defferms{\eqalign{ \f_{kl} &= {1\over 2!} \com{\f_k}{\f_l} \cr
\f_{klm} &= {1\over 3!} ( \f_k \f_l \f_m - \f_k \f_l \f_m + \ldots ) \cr
\f_{klmn} &= {1\over 4!} (\f_k \f_l \f_m \f_n - \f_k \f_l \f_m \f_n + \ldots
)\cr} }
We wish to answer the following question: what are the allowed operators in
\expansion? Note that the six fermion term and the $\com{f_{amnk}}{f_{bmnl}}$
contribution to the two fermion term in \plugin\ automatically vanish at this
order.

To evaluate the terms appearing in \plugin, we can use the following
observation. The right
hand side is an operator $T_{ab}$ symmetric in $a$ and $b$. We can therefore
expand it in
basis elements,
$$ T_{ab} = {1\over 16} \left( \delta_{ab} \delta_{cd} T_{cd} + \g^i_{ab}
\g^i_{cd} T_{cd}
+ {1\over 4!} \g^{ijkl}_{ab} \g^{ijkl}_{cd} T_{cd} \right). $$
An analogous expansion can be performed for antisymmetric operators in terms of
$( \g^{ij}, \g^{ijk})$. The constraints come from requiring that the
coefficients of $(
\g^i, \g^{ijkl} )$ vanish. We can start by evaluating the first term with no
fermions in \plugin,
$$  \left\{ f_{ac}, f_{bc} \right\} = 2 \delta_{ab} ( D^2 + D_i D_i +
D_{ij} D_{ij} ) + 2 \g^i_{ab} \acom{D_j}{ \delta_{ij} D + 2 D_{ij}}. $$
To get a constraint, we need to say something about the three fermion term. Let
us make a few
general remarks. A tensor, $T_{a b_1 \cdots b_{2k+1}} $, antisymmetric in the
$b_j$
indices can be expressed in the following convenient way:
\eqn\AB{T_{ab_1 \cdots b_{2k+1}} = ( B^{I_1}_{a[b_1} A^{I_2}_{b_2 b_3} \cdots
A^{I_{k+1}}_{b_{2k} b_{2k+1} ] }) t_{I_1 I_2 \cdots I_{k+1}}.}
In this expansion, $A \in ( \g^{ij}, \g^{ijk} )$ while $B$ is any basis
element. The index
$I_j$ stands for a collection of $ Spin(9)$ vector indices. In
particular, the term $f_{abcd}$ can be written:
\eqn\threeferm{ f_{abcd} = B^{I}_{a[b} \g^{ij}_{cd]} \, t_{Iij} +
B^{I'}_{a[b} \g^{ijk}_{cd]} \, t'_{I'ijk}.}
Let us demand that our supercharges be invariant under the symmetry CPT which
acts as
complex conjugation and sends,
$$ x \r -x \qquad \qquad p \r p. $$
We can now list the possible structures that can appear in  the first term
of \threeferm,
\eqn\struct{\eqalign{   (1)\quad  & x^j \g^i_{a[b} \g^{ij}_{cd]} \cr
                        (2)\quad & \g^{ij}_{a[b} \g^{ij}_{cd]} \cr
                        (3)\quad & x^k \g^{ijk}_{a[b} \g^{ij}_{cd]}, \cr}}
where we recall that no momenta can appear in \threeferm\ at this order. Using
the
identities in Appendix A, we see that $(2)$ vanishes and that $(1)$ and $(3)$
are
proportional. Similarly, the possible structures for the second term in
\threeferm\
take the form,

\eqn\struct{\eqalign{   (1')\quad  & x^k \g^{ij}_{a[b} \g^{ijk}_{cd]} \cr
                        (2')\quad & \g^{ijk}_{a[b} \g^{ijk}_{cd]} \cr
                        (3')\quad & x^l \g^{ijkl}_{a[b} \g^{ijk}_{cd]}. \cr}}
Again $(2')$ vanishes and $(1')$ and $(3')$ are proportional to $(1)$. So there
is
only one unique structure at this order:
\eqn\threefin{ f_{abcd} = F(r) x^j \g^i_{a[b} \g^{ij}_{cd]}. }
To be hermitian and CPT invariant, the function $F$
appearing in \threefin\ must be imaginary. In evaluating the
contribution of $ \acom{f_{acde}}{f_{bcde}} $ to the coefficient of
$ \g^i_{ab}$, we want to compute $ \g^i_{ab} \acom{f_{acde}}{ f_{bcde}} $. This
involves
a trace of seven gamma matrices which automatically vanishes. This gives our
first
constraint:
\eqn\firstcon{\acom{D_j}{ \delta_{ij} D + 2 D_{ij}} =0, }
for every $i$.

The second possible constraint comes from the coefficient of the
$\g^{ijkl}_{ab}$ term with
 no fermions. There is no contribution from $ \acom{f_{ac}}{f_{bc}} $. For  $
\acom{f_{acde}}
{f_{bcde}} $ to contribute, the trace $ \g^{ijkl}_{ab}\acom{f_{acde}}
{f_{bcde}} $ must be
non-zero. However, a quick inspection of this trace shows that it vanishes;
therefore,
there are no further constraints
from the term with no fermions in \plugin.

 Before examining the term with two fermions
in \plugin, let us expand the $h_k$ in powers
of momenta. We can use the notation $ \h{k}{n,2m}$
to denote a radial function that comes with the following combination of $x$
and $p$:
$ x_{i_1} \cdots x_{i_n} p_{i_1} \cdots
p_{i_n} (p^2)^m $. Then to leading order, we
need to consider:
$$ \eqalign{ D &= \h{1}{0,0} + \h{1}{1,0} ( x \cdot p) \cr
D_i & = \h{2}{0,0} p_i  + \h{3}{1,0} x_i \, ( x \cdot p) + \h{3}{0,0} x_i \cr
D_{ij} & = \h{4}{0,0} (x_i p_j - x_j p_i). \cr}$$
CPT invariance together with hermiticity kills $\h{1}{1,0}$, the real part of
$ \h{3}{0,0}$ and $D_{ij}$. These symmetries further fix the imaginary part of
$\h{3}{0,0}$
 in terms of $\h{2}{0,0}$ and $\h{3}{1,0}$. Our general constraint
\firstcon\ then implies that $D=0$. We are therefore left with,
$$ D_i  = {1\over 2} \left( \h{2}{0,0} p_i  + \h{3}{1,0} x_i \, ( x \cdot p)
+ p_i \h{2}{0,0} + ( p \cdot x) \,  \h{3}{1,0} x_i  \right), $$
with unknown real functions $\h{2}{0,0}, \h{3}{1,0}$. We have finally reduced
the possible
form of the supercharge down to the form we would obtain by canonically
quantizing the
Lagrangian considered in section 2.1. Since the only metric compatible with
supersymmetry
is the flat metric, we can set $ \h{3}{1,0}=0, \h{2}{0,0}=1$ and $F(r)=0$
leaving
$D_i = p_i.$

\subsec{A quick death for terms of order $\hbar^{2}$}

Our supercharge is the free particle charge $Q_a^0$ to lowest order, and we can
now consider deformations which are higher order in $\hbar$. At the next order,
we
can expand our operators in momenta and impose CPT to get:
$$ \eqalign{ D &={1\over 2} \left( \h{1}{2,0} (x\cdot p)^2 + \h{1}{0,2} p^2 +
{\rm h.c.}
\right) \cr
D_i &= p_i \cr
D_{ij} &= \h{4}{1,0} (x\cdot p) (x_i p_j - x_j p_i). \cr
}$$
Our charge is then of the form,
$$ Q = Q^0 + \delta Q, $$
where $ \delta Q $ contains all terms of order at least $\hbar^{2}$. The most
fermionic term has
five fermions with no momenta. The general constraint \firstcon\ still applies
since to
order $\hbar^3$, the purely bosonic part of the Hamiltonian gets no
contributions from
terms with more than one fermion in $Q$. This constraint is easy to analyze,

$$ \eqalign{ {1\over 2}\acom{D_j}{ \delta_{ij} D + 2 D_{ij}}  =& \h{1}{2,0}
x^m x^n p_m p_n p_i +
 \h{1}{0,2} p^2 p_i  \cr & + \h{4}{1,0}(x^m x^i p^m p^2 - x^m x^n p_m p_n p_i)
+ \ldots,}$$
where the omitted terms have fewer powers of momenta. A quick glance tells us
that
for this expression to vanish, $ \h{4}{1,0}=0$ and therefore $\h{1}{2,0} =
\h{1}{0,2}=0$.
To find interesting deformations, we need to go to the next order.

\subsec{A study of terms of order $\hbar^{3}$}

By imposing CPT and hermiticity, we may set $D=D_{ij}=0$ at this order. We are
left with,
\eqn\mixing{ \eqalign{ D_i = &  p_i + {1\over 2} \Bigl( \h{2}{2,0} (\xp )^2 p_i
+
x_i \h{3}{3,0} (\xp )^3 \cr & + \h{2}{0,2} p^2 p_i + x_i \h{3}{1,2} (\xp ) p^2
+
{\rm h.c.} \Bigr).\cr}}
Only two of the four functions appearing in \mixing\ are actually independent.
Gauge
transformations of the form,
\eqn\gaugemix{C = c(r) (\xp )^3 + {\rm h.c.},}
or,
\eqn\secondgauge{C = c(r)(\xp ) p^2 +  {\rm h.c.},}
mix either $ \h{2}{2,0}$ with $\h{3}{3,0}$ or $\h{2}{0,2} $ with $\h{3}{1,2}$.

Since we are considering terms in $H$ of at most order $\hbar^4$, we only need
to consider:
$$ \acom{Q^0}{ \delta Q}. $$
That we do not need to consider terms quadratic in $\delta Q$ will simplify our
computations considerably. The terms in \plugin\ with no fermions give no
constraints
on $D_i$. We need to consider the two fermion terms in \plugin. There are three
non-vanishing terms at this order,
\eqn\twoconstr{  T_{ab}= \com{f_{ak}}{ f_{bl}} + 3 \acom{f_{ac}}{ f_{bckl}}
+ 3 \acom{f_{bc}}{ f_{ackl}},}
but when we trace with either $\g^i_{ab}$ or $ \g^{ijkl}_{ab}$ to find
relations on
the $h$ functions, the last two terms give the
same contribution.

The first term can be written as,
\eqn\firstterm{\com{f_{ak}}{ f_{bl}} = {1\over 16} \bigg( {1\over 4!}
\g^{nmrs}_{ab}
\left\{\g^i\g^{nmrs} \g^j \right\}_{kl} +
\delta_{ab} \g^{ij}_{kl} +
\g^{m}_{ab} \g^{imj}_{kl} \bigg) \com{D_i}{D_j},}
where we take the terms antisymmetric in $k$ and $l$, and where we can replace
$ \com{D_i}{D_j}$ by the general form $ k(x,p) (x^i p^j - x^j p^i)$.

We need to examine the form of $f_{bckl}$ compatible with CPT and hermiticity.
Let us
start by determining the number of independent structures. The first structure
is
the unique vector structure $\g^j_{b[c} \g^{ij}_{kl]}$ with one $Spin(9)$ index
that
already appeared at lowest order in \threefin. This structure can appear with a
number
of CPT invariant functions such as $ F(r) x^i p^2$ where $F$ is imaginary. The
resulting three fermion term is hermitian. Note that this vector structure,
which is
completely antisymmetric in $c,k,l$, can be expressed as a linear combination
of the
four structures,
$$ \left\{ \g^j_{bc} \g^{ij}_{kl}, \quad \g^{ijk}_{bc} \g^{jk}_{kl},
\quad \g^{jk}_{bc} \g^{ijk}_{kl}, \quad \g^{ijkl}_{bc} \g^{jkl}_{kl}
\right\},$$
which are only antisymmetric in $k,l$.

We can also consider tensor structures with
two vector indices; for example, $\g^i_{a[b} \g^{ijk}_{cd]}$, which could
appear with
$ F(r) x^j p^k (\xp )$ where $F$ is now real. However, the resulting three
fermion
term is not hermitian for any of these structures. The last possibility
is a tensor structure with three vector indices. There are four possible
structures,
\eqn\fourstr{ \left\{ \g^i_{a[b} \g^{jk}_{cd]}, \quad \g^{im}_{a[b}
\g^{jkm}_{cd]},
\quad \g^{ijm}_{a[b} \g^{km}_{cd]}, \quad \g^{ijmn}_{a[b} \g^{kmn}_{cd]}
\right\}, }
which are not all independent.  These tensor
structures can give CPT invariant, hermitian three fermion terms so we must
include them.

At this order, the three fermion term can be written in the general form:
\eqn\threegen{\eqalign{ f_{bckl} \f_c \f_k \f_l = &
\{ \,   \g^i_{bc} \g^{jt}_{kl}\, a_{ijt} +
\g^{im}_{bc} \g^{jtm}_{kl} \, b_{ijt} + \cr &
\g^{ijm}_{bc} \g^{tm}_{kl} \, c_{ijt} +
\g^{ijmn}_{bc} \g^{tmn}_{kl} \, d_{ijt}  \, \}
\f_c \f_k \f_l + {\rm h.c.}}}
We have hidden the vector structures in the choice of tensors $a,b,c,d$ and we
have
also only presented an expression manifestly antisymmetric in $k,l$. We can now
compute,

\eqn\secondterm{ \eqalign{ \acom{f_{ac}}{ f_{bckl}} = & {1\over 16} \bigg(
\g^m_{ab} \, \tr( \g^u \g^m \g^{ir}) \g^{jtr}_{kl} \acom{p_u}{b_{ijt}}
+  {1\over 4!} \g^{nmrs}_{ab} \, \tr(\g^u \g^{nmrs} \g^{ijq})\g^{tq}_{kl}
\acom{p_u}{c_{ijt}}
\cr & + {1\over 4!} \g^{nmrs}_{ab} \, \tr(\g^u \g^{nmrs}
\g^{ijqw})\g^{tqw}_{kl}
\acom{p_u}{d_{ijt}}
\bigg) + \ldots,}}
where the omitted terms are either proportional to $\delta_{ab}$ or
antisymmetric in $a,b$.
Note that $a_{ijt}$ does
not appear in \secondterm. Let us first analyze the constraint from $
\g^{nmrs}_{ab}$:

\eqn\firstcon{ \eqalign{\left\{  \g^i\g^{nmrs} \g^j \right\}_{kl} \, k(x,p)
(x^i p^j - x^j p^i) + \cr  6 \, \tr(\g^u \g^{nmrs} \g^{ijq})\g^{tq}_{kl}
\acom{p_u}{c_{ijt}}
+ & \cr  6 \, \tr(\g^u \g^{nmrs} \g^{ijqw})\g^{tqw}_{kl}
\acom{p_u}{d_{ijt}} & =0.}   }
To cancel the piece proportional to $k(x,p)$, the tensor $c$ must contain a
vector term:
$$ c_{ijt} = {1\over 384} k(x,p)( \delta_{it} x^j - \delta_{jt} x^i ) +
\ldots.$$
The other tensor $d$ must contain a vector piece:
$$ d_{ijt} = {1\over 1152} k(x,p)(\delta_{it} x^j - \delta_{jt} x^i ) + \ldots.
$$
We know that the allowed terms in \threegen, after antisymmetrizing the $c,k,l$
indices,
contain a {\it unique} vector structure $ \g^j_{b[c} \g^{ij}_{kl]}$. We can
expand,
$$  12 \g^j_{b[c} \g^{ij}_{kl]} = -7 \g^j_{bc} \g^{ij}_{kl} - \g^{ijt}_{bc}
\g^{jt}_{kl}
+ \g^{jt}_{bc} \g^{ijt}_{kl} - {1\over 6} \g^{ijts}_{bc} \g^{jts}_{kl}, $$
but the ratio of the coefficients of the terms that correspond to the vector
parts
of $ c_{ijt}$ and
$d_{ijt}$ (the second and fourth term, respectively) do not agree with the
solutions for
$c$ and $d$ found above.\foot{That the unique vector structure cannot cancel
the
curvature term proportional to $k(x,p)$ in \firstcon\ can be computed directly
or
confirmed with Mathematica.}
Therefore there is no solution
unless we set $ k(x,p)=0$.  That $k(x,p)=0$ implies that,
$$ \com{p_i}{D_j}=0,$$
to the order of interest. In turn, this implies that the $h$ functions are
precisely of the
form that can be removed by gauge transformations \gaugemix\ and \secondgauge.
We can then
conclude that there are no deformations of the free abelian theory which
generate non-gauge $p^4$ terms. It
would be interesting to see whether this result extends to higher terms in the
derivative expansion.

\newsec{Constraining the Lagrangian}
\subsec{The structure of the supersymmetry transformations}

In this section, we will examine the restrictions that supersymmetry and
$Spin(9)$ invariance
impose on the effective Lagrangian. In section
(2.1), we showed that to order $v^2$ the Lagrangian contains the terms:

\eqn\vsquare{ L_1= \int \, dt \left( {1 \over 2} \, v^2 + i \, \f  \df
\right).}
The supersymmetry tranformations are those given in \transforms\
with $M=0$. At order $v^4$, we must consider all terms,

\eqn\vfourth{ L_2= \int \, dt \left( f_2^{(0)}(r) \, v^4 + \ldots +
f_2^{(8)}(r) \,\f^8
\right), }
which are in the supersymmetric completion of $v^4$.
The terms that we have not written generally contain
accelerations as well as fermions $\f$ with more than a single time derivative.
To
find restrictions on these terms, it is necessary to first understand
the general  form of the supersymmetry transformations.  It is clear that the
free-particle
supersymmetry transformations have to be
modified by higher order terms when \vfourth\ is added to \vsquare, if the
supersymmetry
algebra is to close on-shell; see, for example \rberg. The new transformations
can be
expressed quite generally in the form:

\eqn\newtransforms{ \eqalign{ \dd x^i & = -i \e \g^i \f  + \e  N^i \f\cr
\dd \f_a &= ( \g^i v^i \e )_a + ( M \e )_a.}}

Note that it is impossible to find a solution where either $N^i$
or $M$ vanish when we consider $L=L_1+L_2$ since the algebra will no longer
close. At order
$v^2$, we could set $N^i$ and $M$ to zero because the free-particle
equation of motion,  $\df =0$, is sufficient to ensure closure of the algebra.
Of course,
the fermion equation of motion is considerably more complicated for $L$.
Constructing
$N^i$ and $M$ is a formidable algebraic task. Fortunately, as we shall explain,
we
will not need to know very much more about \newtransforms\ to constrain $f_2$.

The corrections to $ \delta x^i$, encoded in $N^i$, are order $n=2$ where $n$
counts the number of time derivatives plus twice the number of fermions.\foot{
We count $\e$ as order $n=-1/2$.} The
corrections to $\delta \f$, encoded in $M$, are order $n=3$. The variation of
$L$
contains two pieces: the first is order $2$ and automatically vanishes for the
variations \newtransforms. The second piece is order $4$ and gets contributions
from $L_1$ and $L_2$, where we act with the free-particle transformations on
$L_2$
and with the corrections on $L_1$.

Our interest is primarily with the eight fermion term, which has quite magical
properties. This is essentially the `top' form in the supersymmetric completion
of $v^4$ and studying this term (and its higher velocity analogues) is a
natural way
to look for constraints on the Lagrangian. The variation of this term in
\vfourth\
schematically contains two pieces,
\eqn\variationeight{ \dd ( f_2^{(8)}(r) \f^8 )  =   \dd f_2^{(8)}(r)  \,\, \f^8
+
f_2^{(8)}(r) \dd  \f^8. }
The second term contains seven fermions to order $4$ and mixes with the
variation of
$L_1$ and with the variation of the six fermion term in $L_2$. The first term
contains
nine fermions and is quite special. No other term in $L_2$ varies into a nine
fermion
term. Can any term from $L_1$ contain nine fermions? After noting that $M$
contains at
most six fermions and $N^i$ at most four fermions, it is easy to see that the
variation
of $L_1$ cannot contain a nine fermion term. We can now conclude that the nine
fermion
term must vanish by itself. If the metric were not flat, as in the
case with eight supersymmetries, then the corresponding variation of $L_1$
could mix with
the nine fermion term. Even in that case, we would still obtain some equations
that the
eight fermion term would need to obey. However, we will not pursue that case
further
here.

Some dimensional analysis is in order. The coupling in this quantum mechanical
theory,
$g^2$, has mass dimension three. In matrix theory, $g^2 = M_{pl}^6 \RV^3$,
where $\RV$
is the size of the longitudinal direction. For purposes of dimension counting,
the action can be written in the following way:
\eqn\daction{ {1\over g^2} \int \, dt \, \sum_{n} v^{2n} f_n(r). }
In perturbation theory, we can expand each $f_n$ in a power series in the
coupling,
$$ f_n = \left( {1\over r^4}\right)^{n-1} \sum_l C_{nl} \left( {g^2\over r^3}
\right)^l,$$
where $l$ counts the number of loops; see, for example \refs{\rbb, \rper}.
There could
also be non-perturbative corrections to the functions $f_n$. To agree with
classical
long distance supergravity,  $f_1=1$ as we showed in section two. The
coefficient
$C_{21}$ was computed in \rDKPS\ and found to be non-vanishing in agreement
with
supergravity. In \rbb, $C_{22}$ was found to vanish supporting the conjecture
that,
$$ f_2(r) \sim {1\over r^7}.$$
We will show that this conjecture is true non-perturbatively.

There have also been discussions of interactions involving spin dependence
\refs{\rspin, \rBHP}. The latter paper \rBHP\ actually involves a quite
non-trivial
loop computation of the eight fermion term, which gives the following
interesting
result:
\eqn\eightfermion{ \eqalign{ f_2^{(8)}(r) \f^8 &=  -15 { \left(1 \over { 2
r}\right) }^{11}
\left( \f \g^{ij} \f
\,\f \g^{jk} \f \,
\f \g^{lm} \f \, \f \g^{mn} \f \right)  \times \cr
&\left( 2 \delta_{in}  \delta _{kl}-
 { 44 \over{r^2}} \delta_{kl} x_i  x_n + {143 \over{r^4}} x_i x_l \, x_l
\, x_n \right)
 \cr }}
We will prove that this is indeed the only form of the eight fermion
term
compatible with supersymmetry up to an overall numerical factor. This
immediately gives
the desired non-renormalization theorem.

\subsec{The eight fermion term}

To prove that this is the only eight fermion term compatible with $Spin(9)$
and supersymmetry, we first need to prove that the structures that appear in
\eightfermion\ are the only possible structures. Then we will fix the
functional
dependence on $r$ using our observations about the variation of this term.

Since the fermions are real, there are only two possible fundamental building
blocks for fermionic terms: $\f \g^{ij} \f$ and $\f \g^{ijk} \f$.
At order $n=4$, the eight fermion term only depends on $x^i$ and not on $v^i$.
Let
 us then consider all possible terms that can be constructed from
just $\f \g^{ij} \f$:
 \eqn\morestruc{\eqalign{   (1)\quad  & \f \g^{ij} \f \, \f \g^{ij} \f
\,
  \f \g^{kl} \f \, \f \g^{kl} \f   \cr
                        (2)\quad & \f \g^{ij} \f \, \f \g^{jk} \f \,
  \f \g^{kl} \f \, \f \g^{li} \f   \cr
                        (3)\quad &  \f \g^{ij} \f \, \f \g^{jk} \f \,
  \f \g^{kl} \f \, \f \g^{lm} \f \,\, x^i x^m \cr
                        (4)\quad &  \f \g^{ij} \f \, \f \g^{kl} \f \,
  \f \g^{kl} \f \, \f \g^{jm} \f \,\, x^i x^m \cr
                        (5)\quad &  \f \g^{ij} \f \, \f \g^{jk} \f \,
  \f \g^{lm} \f \, \f \g^{mn} \f \,\, x^i x^k x^l x^n. \cr  }}
Using the identities in Appendix A, we can see that (1) and (4) vanish. The
remaining three terms are independent and are actually the three structures
that
appear in \eightfermion.

We should also consider the terms that contain mixed products of $\f
\g^{ij} \f$
and $\f \g^{mnq} \f$. Observe that terms that contain an odd number of
each structure are forbidden by CPT. We are then left with terms of
the form,
\eqn\emorestruc{\eqalign{   (1)\quad  & \f \g^{ijk} \f \, \f \g^{jk} \f
\,
  \f \g^{mn} \f \, \f \g^{imn} \f   \cr
                        (2)\quad &  \f \g^{ij} \f \, \f \g^{ij} \f \,
  \f \g^{lnm} \f \, \f \g^{lmn} \f \cr
                        (3)\quad  & \f \g^{ij} \f \, \f \g^{jk} \f
\,
  \f \g^{kmn} \f \, \f \g^{imn} \f   \cr
                        (4)\quad &  \f \g^{ij} \f \, \f \g^{kl} \f \,
  \f \g^{ljm} \f \, \f \g^{kim} \f \cr
                        (5)\quad &  \f \g^{ij} \f \, \f \g^{kj} \f \,
  \f \g^{lmn} \f \, \f \g^{lmn} \f \,\,  x^i x^k \cr
                        (6)\quad &  \f \g^{ij} \f \, \f \g^{kl} \f \,
  \f \g^{lmn} \f \, \f \g^{imn} \f \,\,  x^i x^k \cr
                        (7)\quad &  \f \g^{ij} \f \, \f \g^{jl} \f \,
  \f \g^{mqn} \f \, \f \g^{lqn} \f \,\,  x^i x^m \cr
                        (8)\quad &  \f \g^{ij} \f \, \f \g^{kl} \f \,
  \f \g^{mjn} \f \, \f \g^{kln} \f \,\,  x^i x^m \cr
                        (9)\quad &  \f \g^{ij} \f \, \f \g^{kl} \f \,
  \f \g^{mkn} \f \, \f \g^{jln} \f \,\,  x^i x^m \cr}}
$$ \eqalign{
                        (10)\quad &  \f \g^{ij} \f \, \f \g^{ij} \f \,
  \f \g^{lmq} \f \, \f \g^{lnq} \f \,\,  x^m x^n \cr
                        (11)\quad &  \f \g^{ij} \f \, \f \g^{jk} \f \,
  \f \g^{kmq} \f \, \f \g^{iqn} \f \,\,  x^m x^n \cr
                        (12)\quad &  \f \g^{ij} \f \, \f \g^{kl} \f \,
  \f \g^{mjk} \f \, \f \g^{nil} \f \,\,  x^m x^n \cr
                        (13)\quad &  \f \g^{ij} \f \, \f \g^{kl} \f \,
  \f \g^{mij} \f \, \f \g^{kln} \f \,\,  x^m x^n \cr
                        (14)\quad &  \f \g^{ij} \f \, \f \g^{jk} \f \,
  \f \g^{klm} \f \, \f \g^{lmn} \f  \,\,x^i x^k x^l x^m  \cr
                        (15)\quad &  \f \g^{ijk} \f \, \f \g^{kl} \f \,
  \f \g^{jmn} \f \, \f \g^{ns} \f  \,\,x^i  x^l x^m x^s. \cr } $$
Each of these structures either vanishes or can be reduced to terms
appearing in \morestruc\ by using the relations in Appendix A.

The last group of terms can be made out of products of $\f \g^{ijk} \f$.
It is not complicated to see that all the nonvanishing structures can be
reduced
again to terms in \morestruc. The most general eight fermion term then takes
the
form:
\eqn\geightf{ \eqalign{ L_{\f^8} = & \big( \f \g^{ij} \f
\,\f \g^{jk} \f \,
\f \g^{lm} \f \, \f \g^{mn} \f \big)
\big( g_1(r) \, \delta_{in}  \delta _{kl}+ \cr &
 g_2(r) \, \delta_{kl} x_i  x_n + g_3(r) x_i x_l \, x_l \, x_n \big). }}

Under a supersymmetry variation,
\eqn\susytr{ \eqalign{ \delta_a \left( f_2^{(8)}\f^8 \right)  =& -i \g^s_{ab}
\f_b
 \left( \f \g^{ij} \f \,\f \g^{jk} \f \,
\f \g^{lm} \f \, \f \g^{mn} \f \right) \times \cr
 & \partial_s \left( g_1(r) \, \delta_{in}
 \delta _{kl}+
 g_2(r) \, \delta_{kl} x_i  x_n + g_3(r) x_i x_k \, x_l \, x_n  \right)+
\ldots,  }}
where the omitted terms either contain seven fermions or are order $6$.
As we pointed out, the term with nine fermions cannot be cancelled by any other
term,
and must vanish by itself. Since \susytr\ is zero, let us apply the operator $
\g^q_{ac}
{ d \over {d \f_c}} \partial_q$ to \susytr,
\eqn\weakcon{\eqalign{   -i\g^q_{ac} { d \over {d \f_c}} \partial_q
\Bigl\{ \g^s_{ab} \f_b \left( \f \g^{ij} \f \,\f \g^{jk} \f \,
\f \g^{lm} \f \, \f \g^{mn} \f \right) \times & \cr
  \partial_s \left( g_1(r) \, \delta_{in}
 \delta _{kl}+
 g_2(r) \, \delta_{kl} x_i  x_n + g_3(r) x_i x_k \, x_l \, x_n  \right)
\Bigr\} & = 0 \cr }}
After summing over the index $a$ this equation gives:
\eqn\condition{\eqalign{  \left( \f \g^{ij} \f \,\f \g^{jk} \f
\,\f \g^{lm} \f \,
 \f \g^{mn} \f \right) \times & \cr
\Delta \left( g_1(r) \, \delta_{in}
 \delta _{kl}+
 g_2(r) \, \delta_{kl} x_i  x_n + g_3(r) x_i x_k \, x_l \, x_n  \right)
 & = \cr
 \left( \f \g^{ij} \f \,\f \g^{jk} \f
\,\f \g^{lm} \f \,
 \f \g^{mn} \f \right)
 \Bigl\{ ({{d^2 g_1} \over {dr^2}}
  +  {8 \over r}{{dg_1} \over {dr}} + 2 g_2)\, \delta_{in}  \delta _{kl}
\cr
  +
 ({{d^2g_2} \over {dr^2}} + {12 \over r} { {dg_2} \over {dr}} + 8 g_3 ) \,
\delta_{kl} x_i  x_n  +
({{d^2 g_3 } \over {dr^2}} +
 {16 \over r} { {dg_3} \over {dr}} ) x_i x_k \, x_l \, x_n \Bigr\} &=0. \cr }}
Since the three  terms appearing in \condition\ are actually independent, we
obtain
three conditions,
\eqn\diffeq{ \eqalign{   {{d^2 g_1} \over {dr^2}} + {8 \over r}{{dg_1}
\over {dr}} + 2 g_2 &= 0  \cr
 {{d^2g_2} \over {dr^2}} + {12 \over r} { {dg_2} \over {dr}} + 8 g_3 &=0
\cr
 {{d^2 g_3 } \over {dr^2}} + {16 \over r} { {dg_3} \over {dr}}  &=0. \cr }}
The solutions to these equations are easily determined,
\eqn\solutions{ \eqalign{
g_1(r)  & = {2 \over 143} { c \over r^{11}} - {1\over 9} {c_1 \over {r^9}} +
{c_4 \over r^7}
+ {2\over 143} c_0 r^4 - {1\over 9} c_2 r^2 + c_3 \cr
g_2(r) & =   -{4\over 13} {c \over r^{13}} +
{c_1 \over r^{11}} -{4 \over 13} c_0
r^2 + c_2 \cr
g_3(r) & = { c \over {r^{15}}} + c_0.  \cr }}
The constraint that we imposed is actually weaker than invariance of the eight
fermion term. Therefore, some of the solutions found in \solutions\ may not
satisfy the stronger invariance condition. On physical grounds, we know that
$c_0, c_2, c_3$ are zero, since the eight fermion term should go to zero
as $ r\r \infty$. The coefficient $c_4$ corresponds to a term that comes with
a negative power of $ g^2$, so we can set it to zero as well. Lastly, $c_1$
corresponds to a term that comes with a positive but fractional power of the
coupling: $g^{2/3}$. This term is clearly not perturbative and we would like to
rule it out.

All these unwanted terms actually correspond to eight fermion terms that do not
satisfy the invariance condition. To see this,  let us apply  $ \g^q_{ac}
{ d \over {d \f_c}} x_q$ to \susytr. This gives three coupled first order
differential
equations which must be satisfied if the eight fermion term is to be
supersymmetric. These equations give stronger constraints than just
harmonicity.
A similar analysis to the one described above shows that we actually need to 
set all the
$c_i$ coefficients in \solutions\ to zero leaving only $c$ non-zero.  The
remaining solution corresponds to the one-loop result computed in \rBHP,
up to an overall numerical factor $c$. The key point is that there are no
higher
loop corrections to the eight fermion term. The only possible non-perturbative
eight
fermion term is given by the solution \solutions.

 This same argument can be extended
to higher velocity terms. In particular, there must exist relations on the
twelve fermion
term in the supersymmetric completion of $v^6$. For the higher velocity terms,
the equations will involve a mixing of the most fermionic term at a given order
with
the variation of
lower order most fermionic terms.  Unravelling these constraints and
studying their implications for M theory should prove exciting.

\bigbreak\bigskip\bigskip\centerline{{\bf Acknowledgements}}\nobreak

It is our pleasure to thank S. de Alwis, R. Entin, O. Ganor, D. Kabat, I.
Klebanov,
B. Ovrut, P. Pouliot, N. Seiberg and E. Witten for helpful
discussions. The work of S.S. is supported by NSF grant
DMS--9627351 and that of M.S. by NSF grant DMS-9505040.  S.P. would like to
thank the Institute for Advanced Study for hospitality during the completion of
this work.

\vfill\eject

\appendix{A}{Fierz Identities}

A collection of Fierz identities used in the text. In this appendix, the
fermions
obey the relation $ \acom{\f_a}{\f_b}=0$ and anticommute with $\e$.

$$\eqalign{\e \g^{ij} \f \,\, \f \g^{ij} \f =& 0 \cr
\e \g^{ijk} \f \,\, \f \g^{ijk} \f =& 0 \cr
\f \g^{ij} \f \,\, \f \g^{ijk} \f =& 0 \cr
\f \g^{i} \f \,\, \f \g^{ij} \f =& 0 \cr
\e \g^{ijk} \f \,\, \f \g^{jk} \f =& 2 \,\e \g^{n} \f \,\, \f \g^{ni} \f
\cr
\e \g^{jkli} \f \,\, \f \g^{jkl} \f =& -6 \,\e \g^{n} \f \,\, \f \g^{ni}
\f \cr
\e \g^{jk} \f \,\, \f \g^{jki} \f =& -2 \,\e \g^{n} \f \,\, \f \g^{ni}
\f \cr
\f \g^{ij} \f \,\, \f \g^{jlk} \f =&  \,\f \g^{ilj} \f \,\, \f \g^{jk}
\f - \f \g^{ln} \f \,\, \f \g^{ink} \f \cr
\f \g^{ipj} \f \,\, \f \g^{jmk} \f =& -3 \,\f \g^{ip} \f \,\, \f \g^{mk}
\f -2 \,\f \g^{im} \f \,\, \f \g^{pk} \f \cr
& + 2  \,\f \g^{ik} \f \,\, \f \g^{pm} \f
+ \delta^{pm} \,\f \g^{ia} \f \,\, \f \g^{ak} \f \cr
&+ \delta^{ik} \,\f \g^{pa} \f \,\, \f \g^{am} \f
- \delta^{im} \,\f \g^{pa} \f \,\, \f \g^{ak} \f \cr
&- \delta^{kp} \,\f \g^{ia} \f \,\, \f \g^{am} \f. \cr
}$$

\listrefs

\end